# Accurate Dose Measurements Using Cherenkov Polarization Imaging.

**Emily Cloutier**[1,2], **Louis Archambault**[1,2], **Luc Beaulieu**[1,2]‡

[1] Physics, physical engineering and optics department and Cancer Research Center, Universite Laval, Quebec, Canada.

[2] CHU de Quebec – Universite Laval, CHU de Quebec, Quebec, Canada

Author to whom correspondence should be addressed. Émily Cloutier : emily.cloutier.1@ulaval.ca

**Abstract.**

**Purpose:** Cherenkov radiation carries the potential of direct in-water dose measurements, but its precision is currently limited by a strong anisotropy. Taking advantage of polarization imaging, this work proposes a new approach for high accuracy Cherenkov dose measurements.

**Methods:** Cherenkov produced in a 15 × 15 × 20 $cm^3$ water tank is imaged with a cooled CCD camera from four polarizer transmission axes [0°, 45°, 90°, 135°]. The water tank is positioned at the isocenter of a 5 × 5 $cm^2$, 6 MV photon beam. Using Malus' law, the polarized portion of the signal is extracted. Corrections are applied to the polarized signal following azimuthal and polar Cherenkov angular distributions extracted from Monte Carlo simulations. Percent depth dose and beam profiles are measured and compared with the prediction from a treatment planning system (TPS).

**Results:** Corrected polarized signals on the central axis reduced deviations at depth from 20% to 0.8 ±1%. For the profile measurement, differences between the corrected polarized signal and the TPS calculations are 1±3% and 8±3% on the central axis and penumbra regions respectively. 29±1% of the Cherenkov signal was found to be polarized.

**Conclusions:** This work proposes a novel polarization imaging approach enabling high precision water-based Cherenkov dose measurements. The method allows correction of the Cherenkov anisotropy within 3% on the beam central axis and in depth.

**Keywords:** Cherenkov, Polarization, Imaging, Luminescence

# 1. Introduction

Passage of electrons in a dielectric medium at a speed higher than the phase speed of light in that medium will lead to the emission of visible light: Cherenkov radiation. Cherenkov radiation has been demonstrated to be useful for real-time monitoring of radiation beams. The use of Cherenkov emission has enabled subsurface tissue dose estimations and real-time treatment delivery verification for breast radiotherapy treatments [1–3]. Cherenkov imaging has also been shown to be a valuable method in radiation beam patterns verification [4, 5]. Cherenkov emission however, occurs only within a cone aligned with the charged particle trajectory. This inherent anisotropy limits the use of Cherenkov imaging for precise dose measurements as the Cherenkov-to-dose proportionality varies according to the source-phantom-detector geometry. In some conditions, the Cherenkov-to-dose ratio can vary up to 20 %, resulting in skewed percent depth dose [6]. Hence, Cherenkov is often treated as a spurious signal that needs to be removed. Still, some have investigated the use of Monte Carlo simulations to calculate Cherenkov-to-dose conversion factors and enable Cherenkov dosimetry [7, 8]. Others have studied the potential of adding a Quinine fluorophore to a water bulk to convert Cherenkov emission into an isotropic scintillation signal or the use of telecentric lenses [9–11].

Despite these challenges, Cherenkov-based dosimetry still carries the potential of perturbation-free, in-water, real-time, and high-resolution dose measurements. Cherenkov emission spectra and intensity are currently well characterized in the context of dose measurements, but little work has been done regarding using its polarization state. A recent work has investigated Cherenkov signal subtraction from fluorescence using a set of two perpendicular polarizers, in the context of carbon ion irradiation [12]. As Cherenkov polarization vector is related to the Cherenkov emission cone and thus its direction [13], our work explores the underlying potential of polarization imaging to resolve Cerenkov anisotropy and achieve direct in-water dose measurements. This letter presents a novel and innovative approach which measures polarization state and angle in order to accurately measure dose based on the Cherenkov signal at any depth or position in a water tank.

**2. Methods**

*2.1. Polarization analysis*

Polarization is a key electromagnetic wave property referring to the electric field oscillation direction. In this work we used a rotating linear polarizer (XP42-18; Edmund Optics, Barrington, NJ) as a filter that transmits light waves of a specific polarization while blocking light waves of other polarization. Measuring a signal from 4 transmission angles [0°, 45°, 90°, 135°] determines valuable information on the intensity and orientation of the light. Using Malus'law [14], along with with the assumption that signal will be partly polarized, the intensity transmitted through a polarizer with a transmission angle of $\alpha_0$ is given by:

$$I = I_{pol} \cdot \cos^2(\alpha_0 - \alpha) + I_{rand.pol.} \qquad (1)$$

The rationale behind this assumption is that many Cerenkov photons will be collected in a single acquisition; the signal is expected to be predominantly emitted at a given polarization angle but contributions from other angles will also be generated as a result of the charged particles angular spread in the phantom. These contributions will balance each other into what is designed as a randomly polarized signal. For each pixel, we fit equation 1 to extract $I_{pol}$, $\alpha$ and $I_{rand.pol.}$, which respectively refer to the polarized contribution of the signal, the mean angle of polarization, and the randomly polarized portion of the signal, using the signal obtained at 0°, 45°, 90° and 135°.

*2.1.1. Cherenkov polarization* Cherenkov light is emitted along a cone whose angle is determined by the velocity of the charged particle. The Cherenkov light direction and production threshold are respectively given by equations 2 and 3:

$$\cos(\theta) = c \cdot [vn]^{-1} \quad (2) \qquad E_{min} = m_0 c^2 \cdot [\sqrt{1 - 1/n^2}]^{-1} \quad (3)$$

where v is the velocity of the charged particle and n the refractive index of the dielectric medium. Cherenkov light is polarized as a result of the electric dipoles oscillations producing a coherent wavefront after the passage of a high-energy electron. The polarization direction coincides with the average spin of that electron [17] and is perpendicular to the cone emission, pointing away from the particle's path [13].

*2.2. Dose measurements*

A 15×15×20 $cm^3$ acrylic tank filled with water was irradiated with a 5×5 $cm^2$, 6 MV photon beam (Clinac iX, Varian Medical Systems, Palo Alto, USA). The water tank surface was aligned at the beam isocenter. A cooled CCD camera (Atik 414EX; Atik Cameras, Norwich, United Kingdom) captured the resulting Cherenkov signal at a distance of 50 cm (figure 1). Measurements were acquired over each polarization axis (0°, 45°, 90° and 135°).

Measurements without a polarizer in front of the camera were also acquired to compare polarized data with raw Cherenkov measurements. The inner walls of the tank were covered with a thin opaque black film to minimize the collection of reflection signal. For each irradiation conditions, ten signal images were acquired and treated with a median temporal filter to reduce transient noise. Ten background frames were averaged and subtracted from the signal images. Frames were further flat field corrected. Room ambient light was minimized by covering the setup with black opaque blankets. Percent depth dose (PDD) and profiles at depth of maximum dose ($d_{max}$) are extracted and compared with dose calculations performed on a treatment planning system (Pinnacle 9.2, Philips Healthcare, Andover, MA), using CT images of the phantom.

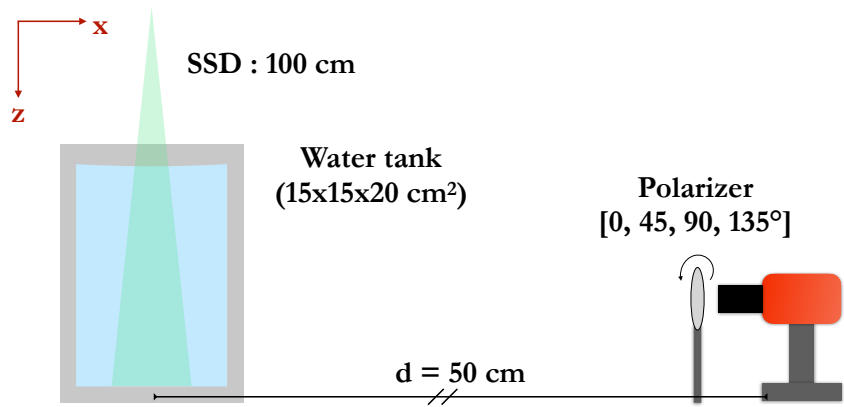

**Figure 1.** Representation of the experimental set-up comprising a 15x15x20 cm³ water tank imaged by a CCD camera from different polarization orientations. The coordinate system is presented in the top left corner.

*2.2.1. Cherenkov dose rectification* We hypothesize that the anisotropic Cherenkov signal, i.e. polarized component, is proportional to the dose, given a correction function taking into account the anticipated directionality of Cherenkov:

$$D(x,y) \propto C_\theta(x,y) \cdot C_\varphi(x,y) \cdot I_{pol}(x,y) \qquad (4)$$

where $C_\theta(x,y)$ and $C_\varphi(x,y)$ refer to the polar and azimuthal angular Cherenkov dependency corrections and $I_{pol}(x,y)$ is found from equation (1). Monte Carlo simulations using the Geant4 toolkit (v4.10.04) [18] were performed to extract the polar and azimuthal distribution of Cherenkov production. A 4×4 cm$^2$ 6 MV phase space from the International Atomic Energy Agency (IAEA) database [19] irradiated a 40×40×40 cm$^4$ water tank (3 × 10$^7$ events). The direction and position of each Cherenkov photon produced were scored. We averaged the $\theta$ and $\phi$ distributions over a 5×5×5 cm$^3$ region-of interest at the center of the irradiation field. To better represent the collection efficiency of the CCD, we collected only the photons having $\theta$>0. These distributions were used to correct skewed Cherenkov dose distributions caused by the intrinsic Cherenkov anisotropy.

## 3. Results and discussion

### 3.1. Polarization analysis

Figure 2 presents a typical regression performed on a pixel dataset obtained from the four transmission polarizer axes. From the regression, it is possible to reconstruct the polarized image, the randomly polarized image, and an image of the angle of linear polarization (AoLP). It would also be possible to calculate the image that would be obtained from any given polarizer transmission axis ($\alpha_0$). For each pixel on the beam central axis, the fits were performed all with $r^2$ > 0.99. These results demonstrate that a set of four polarization measurements (0°, 45°, 90° and 135°) is sufficient to accurately measure the angle and degree of polarization in the context of Cherenkov dose measurements. Hence, a grid-wire sensor, such as the Sony IMX250MZR, would be well suited to provide the necessary information in a single acquisition [20].

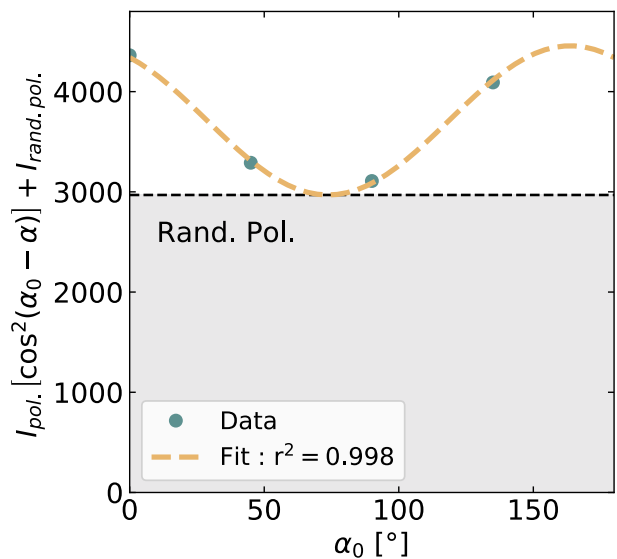

**Figure 2.** A typical example of a data set and associated regression for a given pixel. $\alpha_o$ corresponds to the polarizer transmission axis.

*3.2. Cherenkov rectification*

Figure 3 presents the polar and azimuthal Cherenkov distributions obtained from Monte Carlo calculations. These also corresponds to the dose correction function $C_\varphi(x,y)$ and $C_\theta(x,y)$ applied to account for Cherenkov directionality. The angle is equivalent to the angle from which a measurement is made. In our case, since the camera is located 50 cm from the beam, $\phi$ ranges approximately from 80 to 100° and $\theta$ from 85 to 95°.

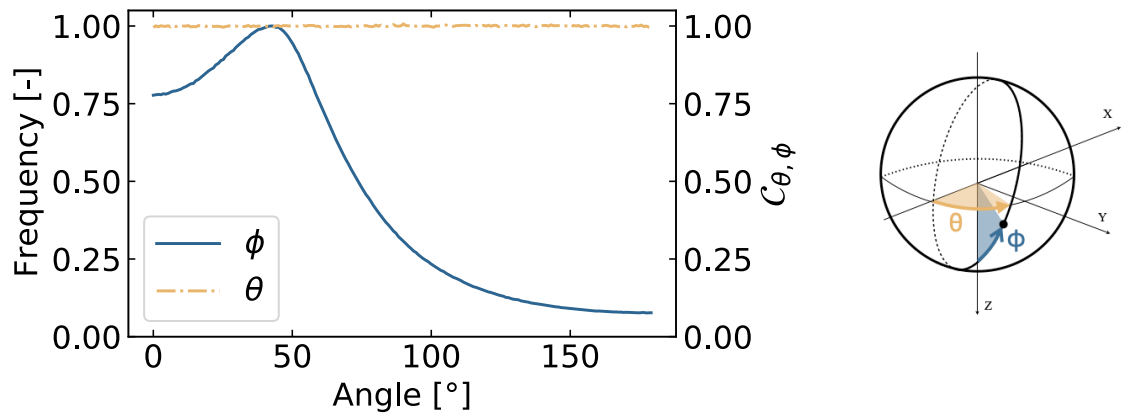

**Figure 3.** Polar and azimuthal Cherenkov distribution produced in water by a 6 MV photon beam. The distributions are measured from Monte Carlo simulations.

*3.3. Dose measurements and corrections*

Figure 4 presents percent depth dose and profiles obtained from polarization analysis. Skewed polarized and randomly polarized measurements are plotted against the raw Cherenkov response obtained from measurements without polarizer, and the TPS dose calculation. The resulting polarized signal corrected for anisotropy using $C_\varphi(x,y)$ and $C_\theta(x,y)$ correction functions is also shown. In depth, the corrected polarized signal present deviations (mean ± standard deviation) of 0.8±1% from TPS calculations. In comparison, the raw Cherenkov signal presents deviations up to 20%.

Comparing polarized corrected Cherenkov signals to the results obtained using telecentric lenses [11], we notice improvements while using a polarization imaging formalism. Telecentric lenses resulted in greater deviations (up to 3%). Additionally, telecentric lenses are limited to a small field of view or are otherwise costly, which limits the convenience of that method.

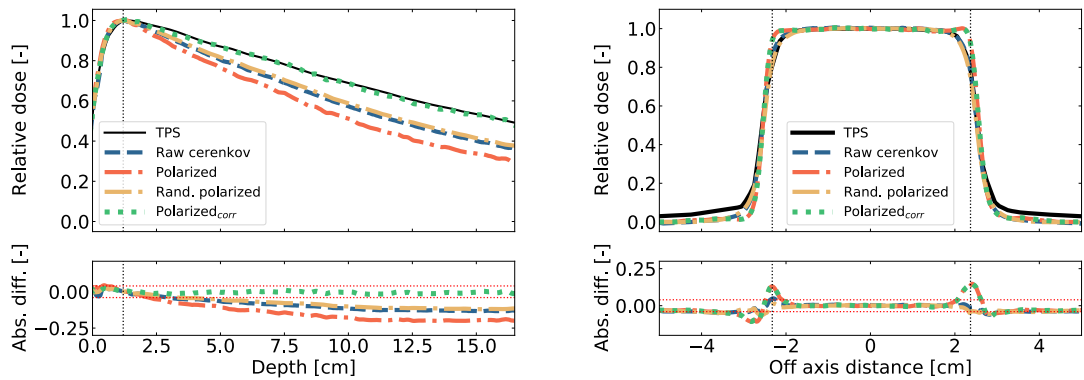

**Figure 4.** Percent depth dose (left) and profiles (right) summed over the thickness of the tank and compared to the prediction from the treatment planning system (TPS). Raw Cherenkov refers to the signal obtained without polarizer in front of the camera. The vertical dotted line indicates the position of maximum after which electronic equilibrium is reached. Horizontal red dotted lines indicate a ±4% difference region.

As for the profiles, the corrected signal and TPS calculations present greater discrepancies in the penumbra region where electronic equilibrium is lost. As such, a higher proportion of the signal is polarized, resulting in a shoulder-shaped excess of signal. The absolute difference between the corrected polarized signal and the TPS calculation are of 1±3% and 8±3% on the central axis and penumbra regions respectively. Regarding the field sizes, the corrected signal results in a field size of 5.13 ± 0.03 cm which compares to the 5.1 ± 0.1 predicted from the TPS. For profile measurements, better agreement is found using the randomly polarized signal.

Overall, 29 ± 1% of the Cherenkov signal was found to be polarized. The resulting 2D images correspond to the sum of Cherenkov produced over the optical axis (or tank thickness). Hence, Cherenkov light from different polarization angles will be summed, resulting in an overall partly polarized signal despite the polarized nature of Cherenkov. Moreover, fluorescence, the water scintillation signal, may also contribute to the randomly polarized signal [21].

## 4. Conclusion

A novel approach was developed using polarization imaging which enabled precise in-water dose measurements based on the Cherenkov signal. Using a set of 4 measurements performed from 4 polarization transmission axes, we were able to measure the polarized and randomly polarized contributions of the Cherenkov signal, as well as the mean angle of linear polarization. Taking advantage of the known Cherenkov angular distribution, we further propose a correction technique that solves the anisotropy issue that otherwise limits the dose measurement precision. Agreements within 2% on the depth dose curve were obtained in a situation which previously presented deviations up to 20%. Agreement is reduced in the beam penumbra region for the profile measurement, but remained acceptable on the central axis, where an average agreement within 3% was obtained. Given the demonstrated benefits of polarization imaging for dose measurements, future work will investigate extending the developed formalism to other photon beam energies as well as electron beam dose measurements. The results presented in this letter open the way to complete perturbation-free in-water Cherenkov 2D dose measurements.

## 5. Acknowledgements

This work was financed by the Natural Sciences and Engineering Research Council of Canada (NSERC) Discovery grants #2019-05038 and #2018-04055. Emily Cloutier acknowledged support by the Fonds de Recherche du Quebec – Nature et Technologies (FRQNT). We thank Ronan Lefol for the English revision of the paper.